\documentstyle[twoside,psfig,amsmath]{article}

\catcode`\@=11
\long\def\@makefntext#1{
\protect\noindent \hbox to 3.2pt {\hskip-.9pt
$^{{\eightrm\@thefnmark}}$\hfil}#1\hfill}       

\def\@makefnmark{\hbox to 0pt{$^{\@thefnmark}$\hss}}    

\def\ps@myheadings{\let\@mkboth\@gobbletwo
\def\@oddhead{\hbox{}
\rightmark\hfil\eightrm\thepage}
\def\@oddfoot{}\def\@evenhead{\eightrm\thepage\hfil
\leftmark\hbox{}}\def\@evenfoot{}
\def\sectionmark##1{}\def\subsectionmark##1{}}

\oddsidemargin=\evensidemargin
\newcounter{sectionc}\newcounter{subsectionc}\newcounter{subsubsectionc}
\renewcommand{\section}[1] {\vspace{12pt}\addtocounter{sectionc}{1}
\setcounter{subsectionc}{0}\setcounter{subsubsectionc}{0}\noindent
    {\tenbf\thesectionc. #1}\par\vspace{5pt}}
\renewcommand{\subsection}[1] {\vspace{12pt}\addtocounter{subsectionc}{1}
    \setcounter{subsubsectionc}{0}\noindent
    {\bf\thesectionc.\thesubsectionc. {\kern1pt \bfit #1}}\par\vspace{5pt}}
\renewcommand{\subsubsection}[1] {\vspace{12pt}\addtocounter{subsubsectionc}{1}
    \noindent{\tenrm\thesectionc.\thesubsectionc.\thesubsubsectionc.
    {\kern1pt \tenit #1}}\par\vspace{5pt}}
\newcommand{\nonumsection}[1] {\vspace{12pt}\noindent{\tenbf #1}
    \par\vspace{5pt}}

\newcounter{appendixc}
\newcounter{subappendixc}[appendixc]
\newcounter{subsubappendixc}[subappendixc]
\renewcommand{\thesubappendixc}{\Alph{appendixc}.\arabic{subappendixc}}
\renewcommand{\thesubsubappendixc}
    {\Alph{appendixc}.\arabic{subappendixc}.\arabic{subsubappendixc}}

\renewcommand{\appendix}[1] {\vspace{12pt}
        \refstepcounter{appendixc}
        \setcounter{figure}{0}
        \setcounter{table}{0}
        \setcounter{lemma}{0}
        \setcounter{theorem}{0}
        \setcounter{corollary}{0}
        \setcounter{definition}{0}
        \setcounter{equation}{0}
        \renewcommand{\thefigure}{\Alph{appendixc}.\arabic{figure}}
        \renewcommand{\thetable}{\Alph{appendixc}.\arabic{table}}
        \renewcommand{\theappendixc}{\Alph{appendixc}}
        \renewcommand{\thelemma}{\Alph{appendixc}.\arabic{lemma}}
        \renewcommand{\thetheorem}{\Alph{appendixc}.\arabic{theorem}}
        \renewcommand{\thedefinition}{\Alph{appendixc}.\arabic{definition}}
        \renewcommand{\thecorollary}{\Alph{appendixc}.\arabic{corollary}}
        \noindent{\tenbf Appendix \theappendixc #1}\par\vspace{5pt}}
\newcommand{\subappendix}[1] {\vspace{12pt}
        \refstepcounter{subappendixc}
        \noindent{\bf Appendix \thesubappendixc. {\kern1pt \bfit #1}}
    \par\vspace{5pt}}
\newcommand{\subsubappendix}[1] {\vspace{12pt}
        \refstepcounter{subsubappendixc}
        \noindent{\rm Appendix \thesubsubappendixc. {\kern1pt \tenit #1}}
    \par\vspace{5pt}}

\newcommand{\textlineskip}{\baselineskip=13pt}
\newcommand{\smalllineskip}{\baselineskip=10pt}

\def\eightcirc{
\begin{picture}(0,0)
\put(4.4,1.8){\circle{6.5}}
\end{picture}}
\def\eightcopyright{\eightcirc\kern2.7pt\hbox{\eightrm c}}

\newcommand{\copyrightheading}[1]
    {\vspace*{-2.5cm}\smalllineskip{\flushleft
    {\footnotesize International Journal of Modern Physics C #1}\\
    {\footnotesize $\eightcopyright$\, World Scientific Publishing
     Company}\\
     }}

\newcommand{\publisher}[2]{{\begin{center}\footnotesize\smalllineskip
    Received #1\\
    Revised #2
    \end{center}
    }}

\def\abstracts#1#2#3{{
    \centering{\begin{minipage}{4.5in}\footnotesize\baselineskip=10pt
    \parindent=0pt #1\par
    \parindent=15pt #2\par
    \parindent=15pt #3
    \end{minipage}}\par}}

\def\keywords#1{{
    \centering{\begin{minipage}{4.5in}\footnotesize\baselineskip=10pt
    {\footnotesize\it Keywords}\/: #1
    \end{minipage}}\par}}

\newcommand{\bibit}{\nineit}
\newcommand{\bibbf}{\ninebf}
\renewenvironment{thebibliography}[1]
        {\frenchspacing
     \ninerm\baselineskip=11pt
         \begin{list}{\arabic{enumi}.}
        {\usecounter{enumi}\setlength{\parsep}{0pt}
     \setlength{\leftmargin 12.7pt}{\rightmargin 0pt} 
         \setlength{\itemsep}{0pt} \settowidth
    {\labelwidth}{#1.}\sloppy}}{\end{list}}

\newcounter{itemlistc}
\newcounter{romanlistc}
\newcounter{alphlistc}
\newcounter{arabiclistc}

\newcommand{\fcaption}[1]{
        \refstepcounter{figure}
        \setbox\@tempboxa = \hbox{\footnotesize Fig.~\thefigure. #1}
        \ifdim \wd\@tempboxa > 5in
           {\begin{center}
        \parbox{5in}{\footnotesize\smalllineskip Fig.~\thefigure. #1}
            \end{center}}
        \else
             {\begin{center}
             {\footnotesize Fig.~\thefigure. #1}
              \end{center}}
        \fi}

\newcommand{\tcaption}[1]{
        \refstepcounter{table}
        \setbox\@tempboxa = \hbox{\footnotesize Table~\thetable. #1}
        \ifdim \wd\@tempboxa > 5in
           {\begin{center}
        \parbox{5in}{\footnotesize\smalllineskip Table~\thetable. #1}
            \end{center}}
        \else
             {\begin{center}
             {\footnotesize Table~\thetable. #1}
              \end{center}}
        \fi}

\def\@citex[#1]#2{\if@filesw\immediate\write\@auxout
    {\string\citation{#2}}\fi
\def\@citea{}\@cite{\@for\@citeb:=#2\do
    {\@citea\def\@citea{,}\@ifundefined
    {b@\@citeb}{{\bf ?}\@warning
    {Citation `\@citeb' on page \thepage \space undefined}}
    {\csname b@\@citeb\endcsname}}}{#1}}

\newif\if@cghi
\def\cite{\@cghitrue\@ifnextchar [{\@tempswatrue
    \@citex}{\@tempswafalse\@citex[]}}
\def\citelow{\@cghifalse\@ifnextchar [{\@tempswatrue
    \@citex}{\@tempswafalse\@citex[]}}
\def\@cite#1#2{{$\null^{#1}$\if@tempswa\typeout
    {IJCGA warning: optional citation argument
    ignored: `#2'} \fi}}

\def\pmb#1{\setbox0=\hbox{#1}
    \kern-.025em\copy0\kern-\wd0
    \kern.05em\copy0\kern-\wd0
    \kern-.025em\raise.0433em\box0}

\def\fnt#1#2{\footnotetext{\kern-.3em
    {$^{\mbox{\scriptsize #1}}$}{#2}}}

\def\ps@myheadings{%
    \let\@oddfoot\@empty\let\@evenfoot\@empty
    \def\@evenhead{\slshape\leftmark\hfil}
    \def\@oddhead{\hfil{\slshape\rightmark}}
    \let\@mkboth\@gobbletwo
    \let\sectionmark\@gobble
    \let\subsectionmark\@gobble
    }
\font\tenrm=cmr10
\font\tenit=cmti10
\font\tenbf=cmbx10
\font\bfit=cmbxti10 at 10pt
\font\ninerm=cmr9
\font\nineit=cmti9
\font\ninebf=cmbx9
\font\eightrm=cmr8

\def\qed{\hbox{${\vcenter{\vbox{            
   \hrule height 0.4pt\hbox{\vrule width 0.4pt height 6pt
   \kern5pt\vrule width 0.4pt}\hrule height 0.4pt}}}$}}

\def\bsc{{\sc a\kern-6.4pt\sc a\kern-6.4pt\sc a}}   
\def\bflatex{\bf L\kern-.30em\raise.3ex\hbox{\bsc}\kern-.14em
T\kern-.1667em\lower.7ex\hbox{E}\kern-.125em X}

\begin{document}
\setlength{\textheight}{7.7truein}  

\thispagestyle{empty}

\markboth{\protect{\footnotesize\it Yu.Yu. Tarasevich \& E.N. Manzhosova}}
{\protect{\footnotesize\it On Site Percolation on Correlated Simple Cubic Lattice}}

\normalsize\textlineskip

\setcounter{page}{1}

\copyrightheading{}         

\vspace*{0.88truein}

\centerline{\bf ON  SITE PERCOLATION }
\vspace*{0.035truein}
\centerline{\bf ON THE CORRELATED SIMPLE CUBIC LATTICE}
\vspace*{0.37truein}
\centerline{\footnotesize YURIY YU. TARASEVICH and ELENA N. MANZHOSOVA }
\baselineskip=12pt
\centerline{\footnotesize\it Institute for Physics and Mathematics, Astrakhan State University,
Tatishchev 20a}
\baselineskip=10pt
\centerline{\footnotesize\it Astrakhan, 414056,
Russia}
\centerline{\footnotesize\it E-mail: tarasevich@astranet.ru}

\vspace*{10pt}      
\publisher{(received date)}{(revised date)}

\vspace*{0.25truein} \abstracts{We consider site percolation on a
correlated bi-colored simple cubic lattice. The correlated medium
is constructed from a strongly alternating bi-colored simple cubic
lattice due to anti-site disordering. The percolation threshold is
estimated. The cluster size distribution is obtained. A possible
application to the double 1:1 perovskites is discussed.}{}{}

\vspace*{5pt} \keywords{Correlated percolation; Percolation
threshold; Simple cubic lattice; Double perovskites.}


\vspace*{1pt}\textlineskip  
\section{Introduction}      
\vspace*{-0.5pt} \noindent In percolation theory one aims to model
random (disordered) media to evaluate their typical
properties.\cite{Stauffer} The simplest example is called
Bernoulli percolation, where all sites have the same probability
to be occupied independent of each other. The assumption of
stochastic independence is too strong for consider many natural
systems. For such systems a medium is not Bernoulli rather then
correlated. Recently, a particular lattice topology and a
particular constraint were investigated.\cite{Reimann,Bendisch} We
will deal with another particular correlated medium.

\begin{figure}[htbp]
\vspace*{13pt}
\centerline{\psfig{file=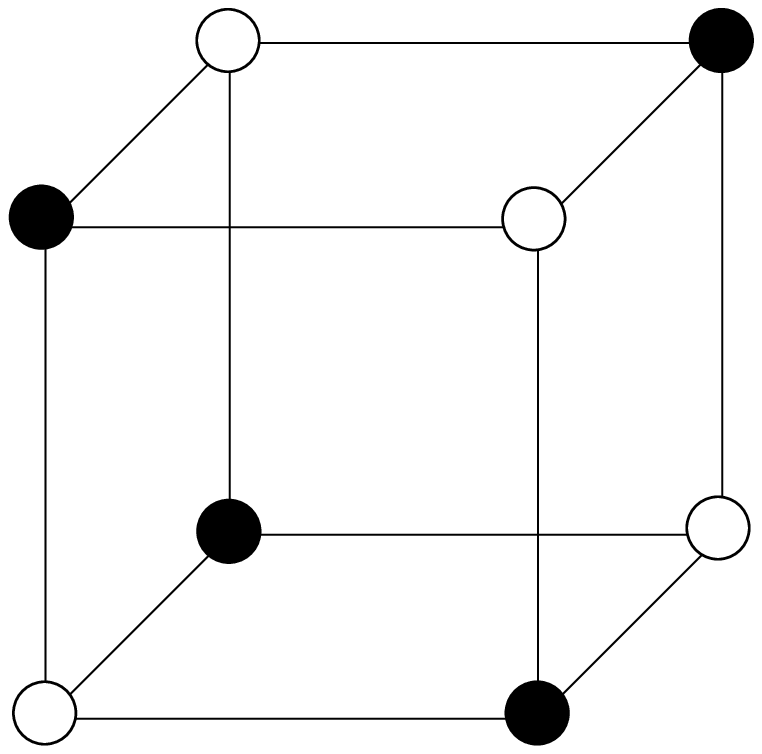,width=0.5\textwidth}} 
\vspace*{13pt} \fcaption{Bi-colored alternating simple cubic
lattice (rock salt lattice).}\label{fig:ascl}
\end{figure}

Let us consider a strongly alternating bi-colored simple cubic
lattice (Fig.~\ref{fig:ascl}). Each `white' site is surrounded by
`black' ones and vice versa, i.e. each of the sites is isolated
and forms a cluster of unit size. The medium is fully ordered. One
can divide the lattice into two sublattices: `black' and `white'.
The probability that a `white' site belongs to the `white'
sublattice is unity. Let us assume that alternating order is
partially destroyed. We will denote $p$ the probability to find a
`white' site in the `black' sublattice. If $p>0$, then one can
find a `white' cluster of size more than the unity. If $p=0.5$
(fully disordered system) then  there is an infinity cluster
indeed, because the site percolation threshold for the simple
cubic lattice and Bernoulli percolation is
$0.311608$.\cite{lorenz_98b} We will look for a critical
disturbance $p_c$ where an infinity cluster is formed, for the
first time.

\section{Cluster size distribution}
\noindent
The probabilities per cluster site $n_s$ that a randomly chosen lattice site belongs
to a cluster of $s$ sites are
\begin{equation}
n_1 =  \frac{1}{2}\left( {(1 - p)^7 + p^7} \right),
\label{eq:n1}
\end{equation}
\begin{equation}
n_2 = 3 p^6(1 - p)^6,
\label{eq:n2}
\end{equation}
\begin{equation}
n_3 = \frac {3}{2} p^5(1-p)^5\left(p^7+(1-p)^7+4p^6+4(1-p)^6\right),
\label{eq:n3}
\end{equation}
\begin{multline}
n_4 = 3p^{11}(1-p)^{11}+ 51p^{10}(1-p)^{10}\\
+6p^4(1-p)^4\left(p^{12}+(1-p)^{12}\right)+4p^4(1-p)^4\left(p^{11}+(1-p)^{11}\right),
\label{eq:n4}
\end{multline}
\begin{multline}
n_5 = \frac{3}{2}
p^{10}(1-p)^{10}\left(p^7+(1-p)^7+8p^6+8(1-p)^6\right.\\ \left.
+32p^5+32(1-p)^5+8p^4+8(1-p)^4\right)\\
+12p^9(1-p)^9\left(p^8+(1-p)^8+3p^7+3(1-p)^7+8\left(p^6+(1-p)^6\right)\right)\\
+\frac{15}{2}p^3(1-p)^3\left(p^{17}+(1-p)^{17}\right).
\label{eq:n5}
\end{multline}

\begin{figure}[htbp]
\vspace*{13pt}
\centerline{\psfig{file=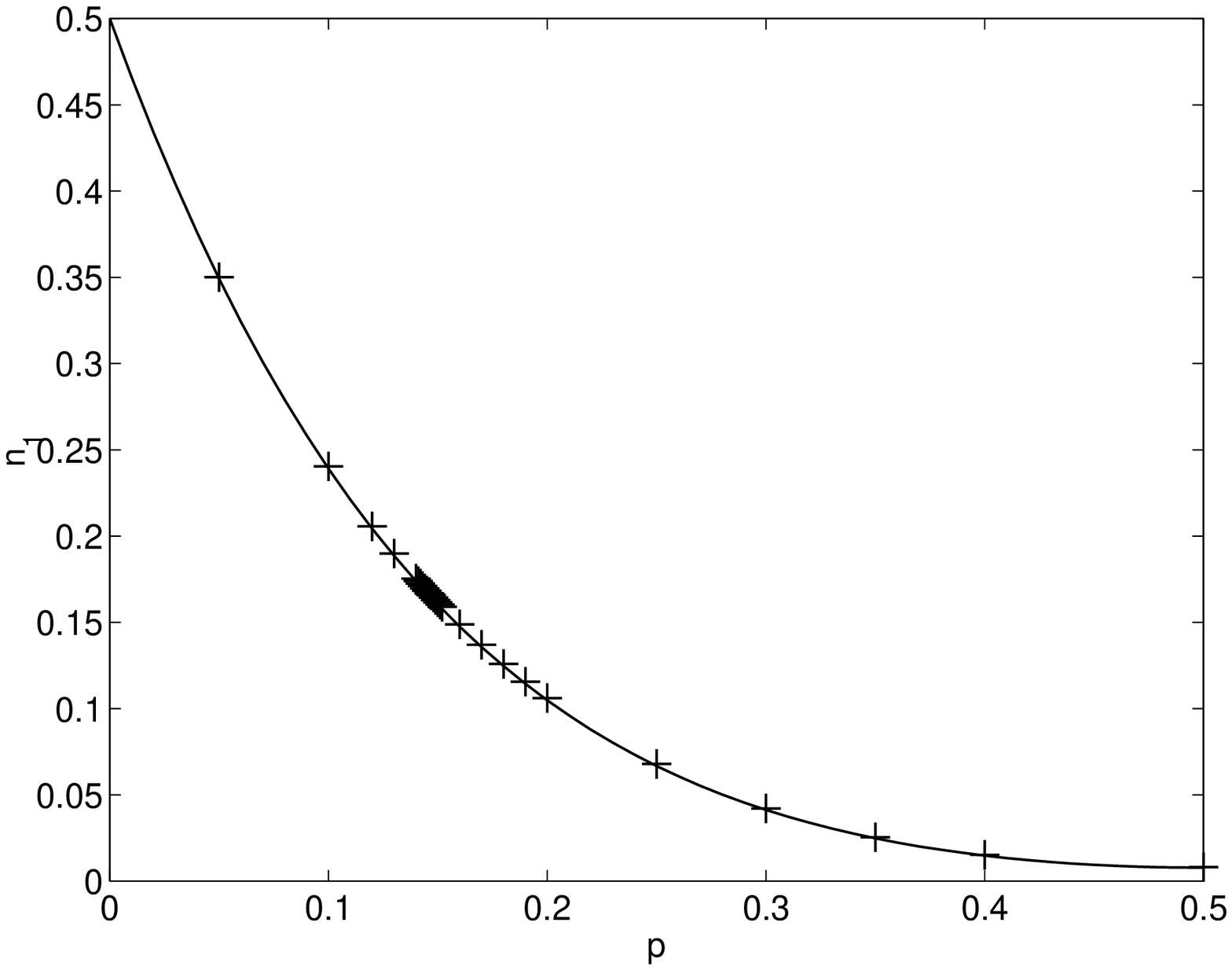,width=0.7\textwidth}} 
\vspace*{13pt} \fcaption{The probability that an arbitrary site
belongs to cluster of size $s=1$. Solid line~--- from
Eq.~\ref{eq:n1}, crosses~--- numerical data. Here and below only
left part of the plot is presented because of symmetry.
}\label{fig:n1}
\end{figure}

\begin{figure}[htbp]
\vspace*{13pt}
\centerline{\psfig{file=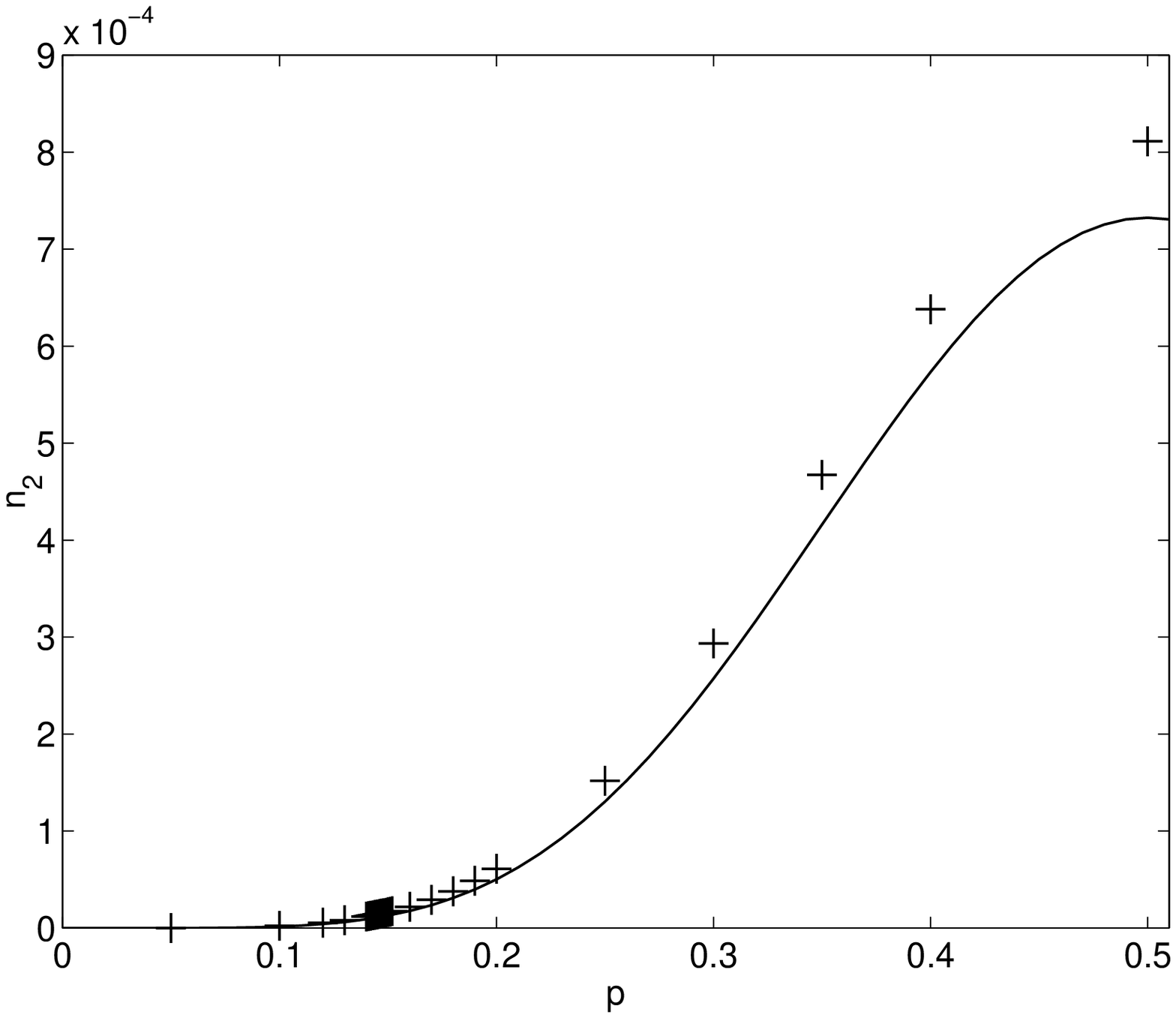,width=0.7\textwidth}} 
\vspace*{13pt} \fcaption{The probability that an arbitrary site
belongs to cluster of size $s=2$. Solid line~--- from
Eq.~\ref{eq:n2}, crosses~--- numerical data. Here and below one
can see a visible deviation from the theoretical curve because of
well known finite size effect for the large
clusters.\cite{Stauffer}}\label{fig:n2}
\end{figure}

\begin{figure}[htbp]
\vspace*{13pt}
\centerline{\psfig{file=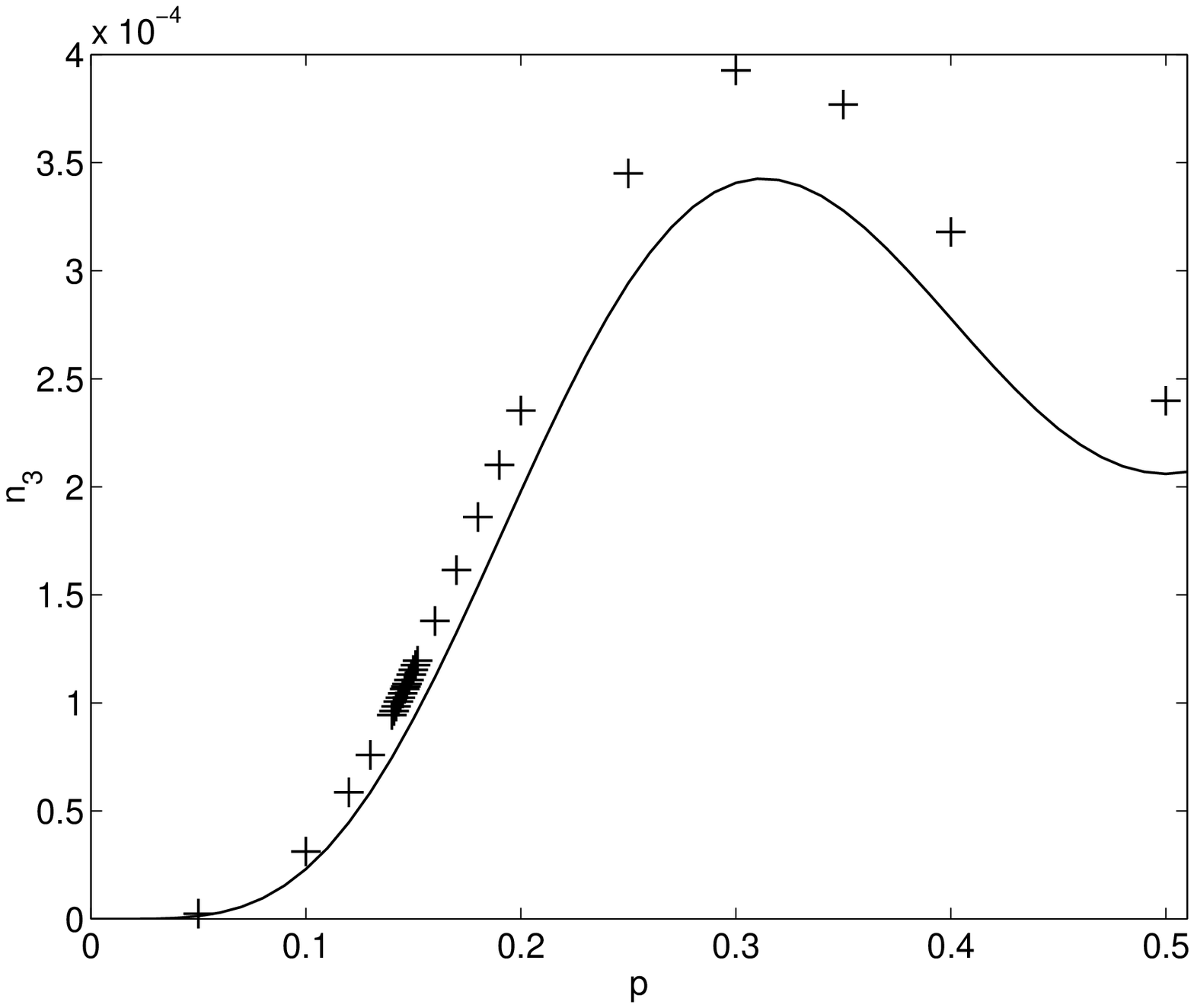,width=0.7\textwidth}} 
\vspace*{13pt} \fcaption{The probability that an arbitrary site
belongs to cluster of size $s=3$. Solid line~--- from
Eq.~\ref{eq:n3}, crosses~--- numerical data. }\label{fig:n3}
\end{figure}

\begin{figure}[htbp]
\vspace*{13pt}
\centerline{\psfig{file=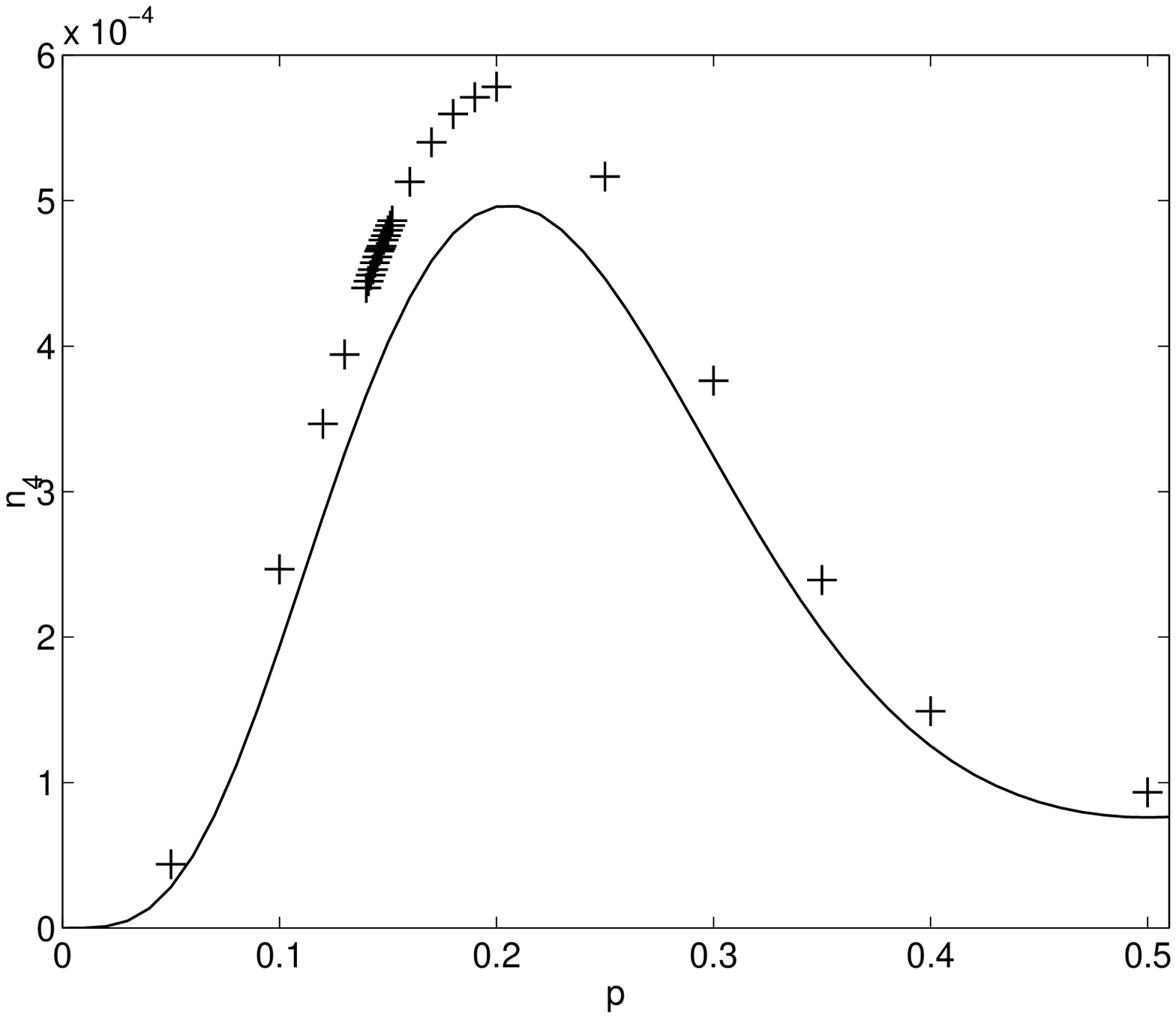,width=0.7\textwidth}} 
\vspace*{13pt} \fcaption{The probability that an arbitrary site
belongs to cluster of size $s=4$. Solid line~--- from
Eq.~\ref{eq:n4}, crosses~--- numerical data. }\label{fig:n4}
\end{figure}

\begin{figure}[htbp]
\vspace*{13pt}
\centerline{\psfig{file=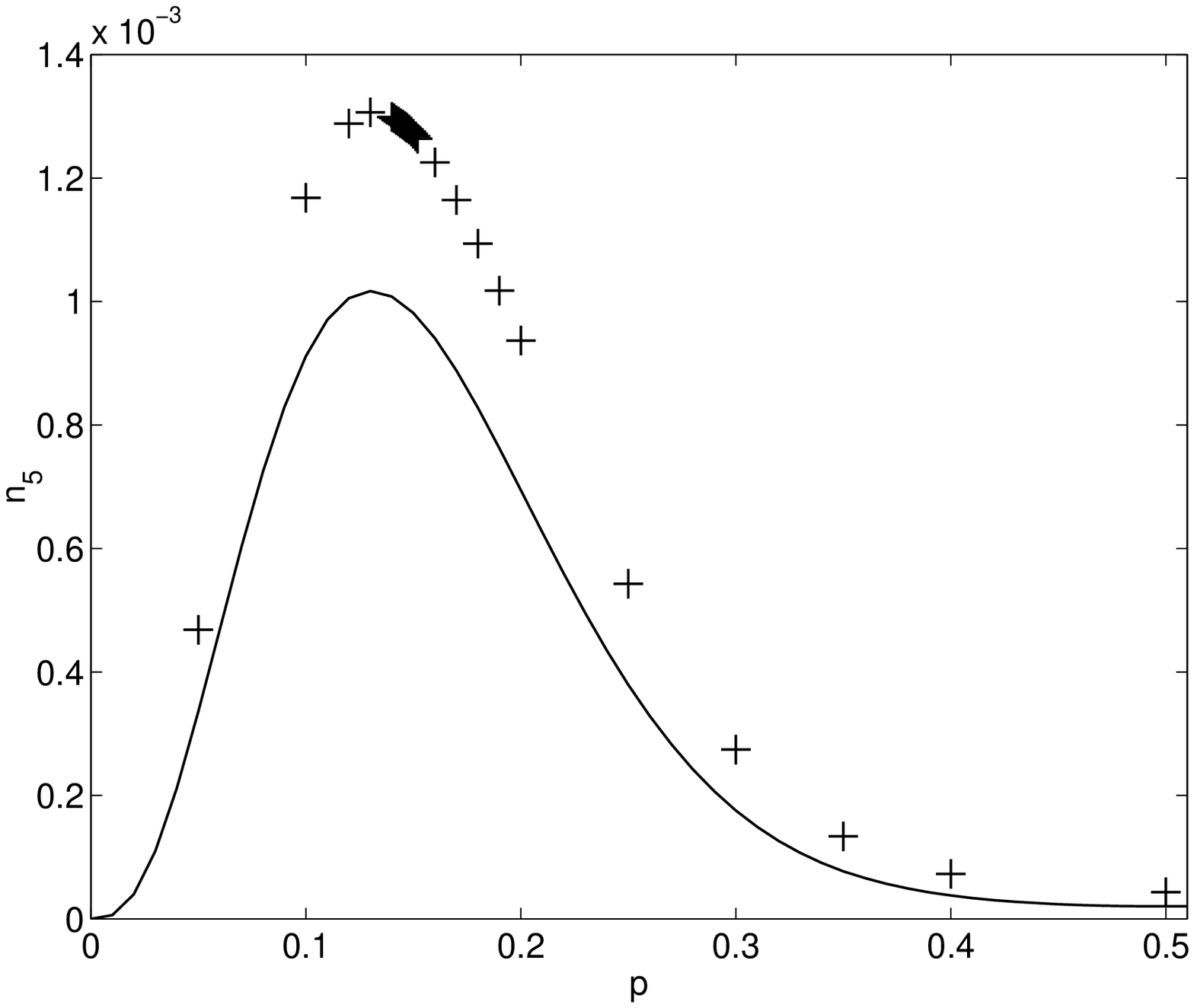,width=0.7\textwidth}} 
\vspace*{13pt} \fcaption{The probability that an arbitrary site
belongs to cluster of size $s=5$. Solid line~--- from
Eq.~\ref{eq:n5}, crosses~--- numerical data. }\label{fig:n5}
\end{figure}

The numerical results were obtained using Hoshen--Kopelman
algorithm.\cite{Hoshen76} We investigated a number of sample
lattices with linear size $L$ up to 128 sites. The theoretical
results and the computer simulations are in the reasonable
agreement (see Fig.~\ref{fig:n1}--\ref{fig:n5}).

\begin{figure}[htbp]
\vspace*{13pt}
\centerline{\psfig{file=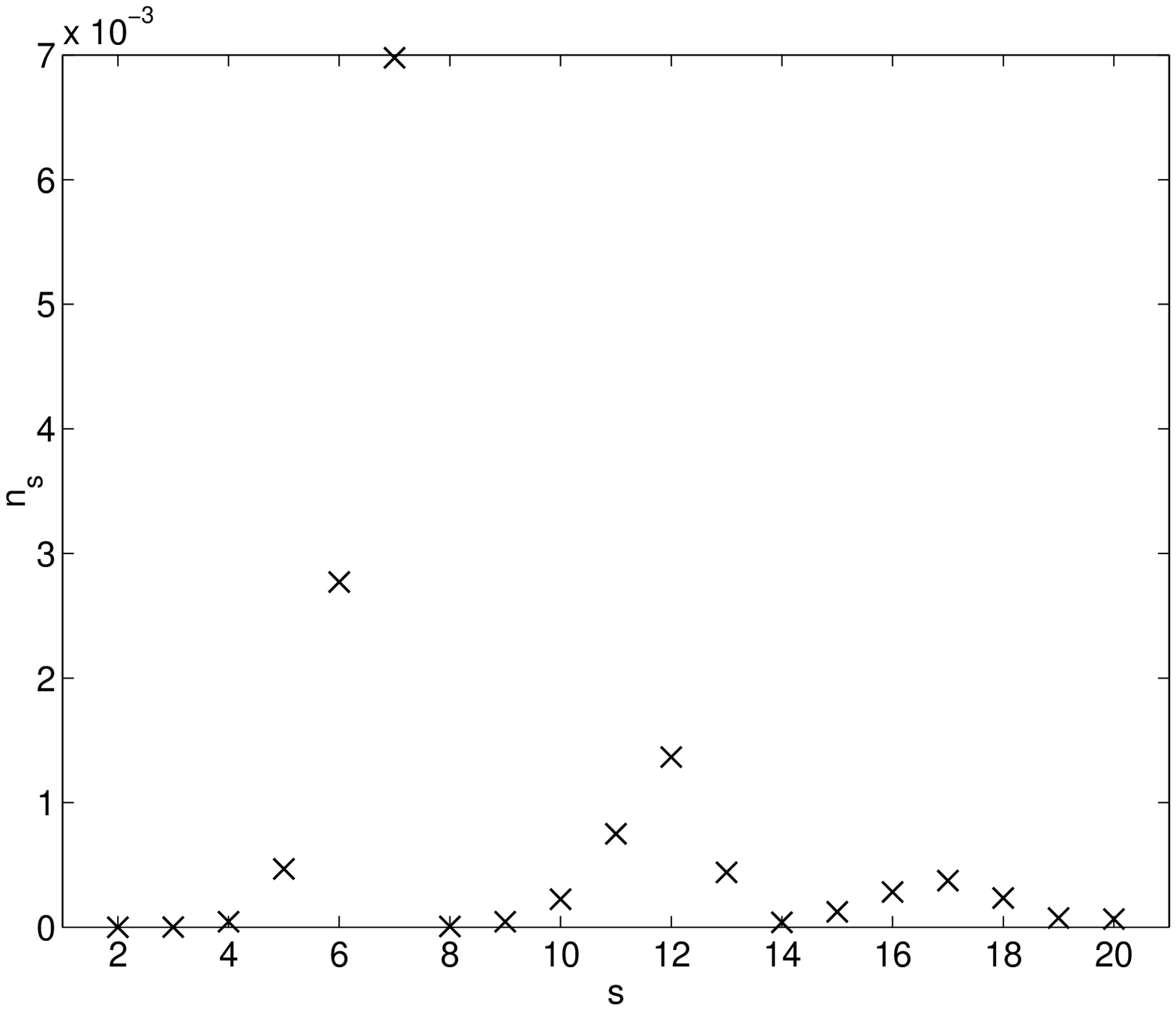,width=0.7\textwidth}} 
\vspace*{13pt} \fcaption{Cluster size distribution for $p=0.05$.
Here and below cluster of unit size is omitted because of very
large magnitude.}\label{fig:s1}
\end{figure}

\begin{figure}[htbp]
\vspace*{13pt}
\centerline{\psfig{file=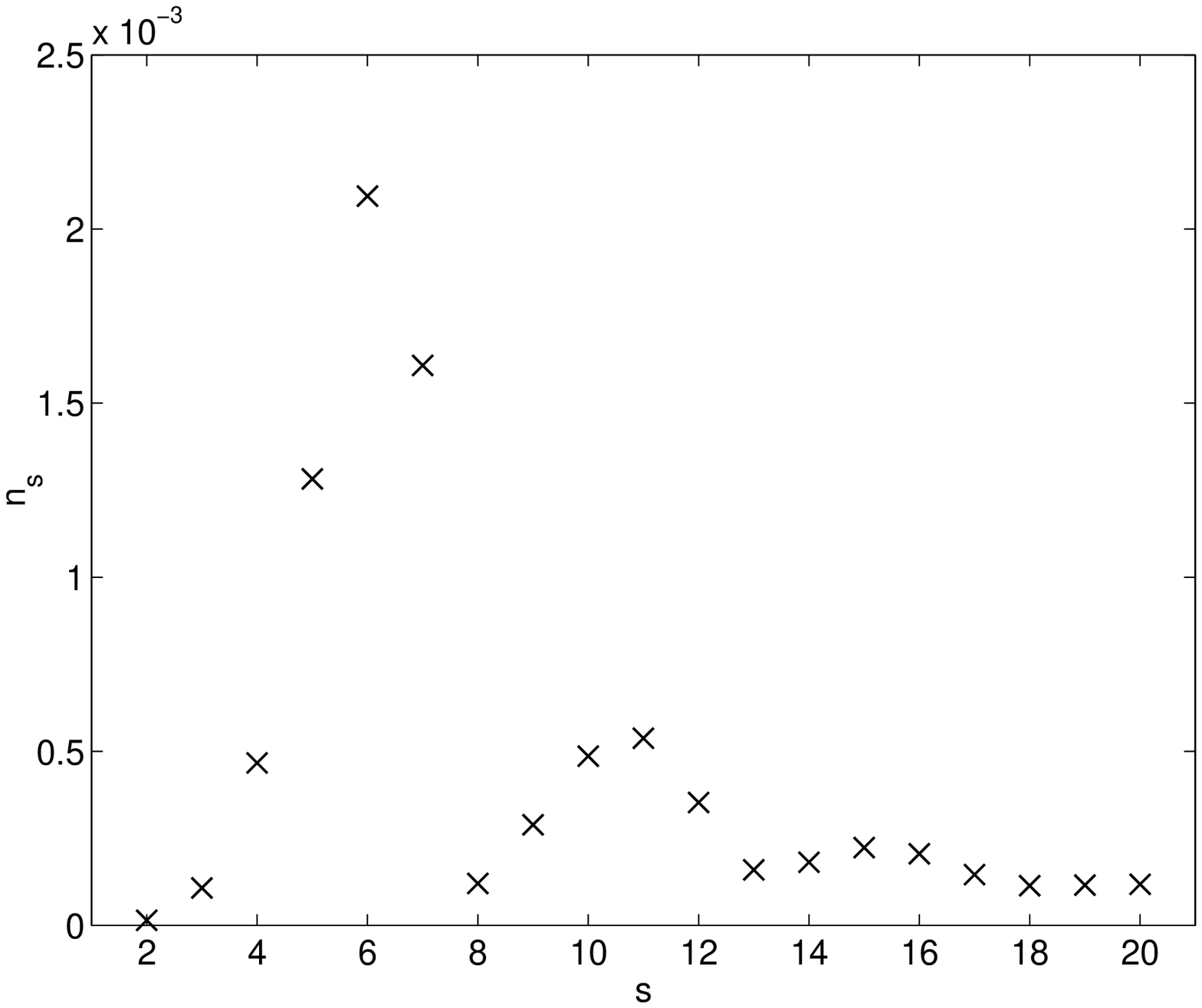,width=0.7\textwidth}} 
\vspace*{13pt} \fcaption{Cluster size distribution for
$p=0.1465$.}\label{fig:s2}
\end{figure}

\begin{figure}[htbp]
\vspace*{13pt}
\centerline{\psfig{file=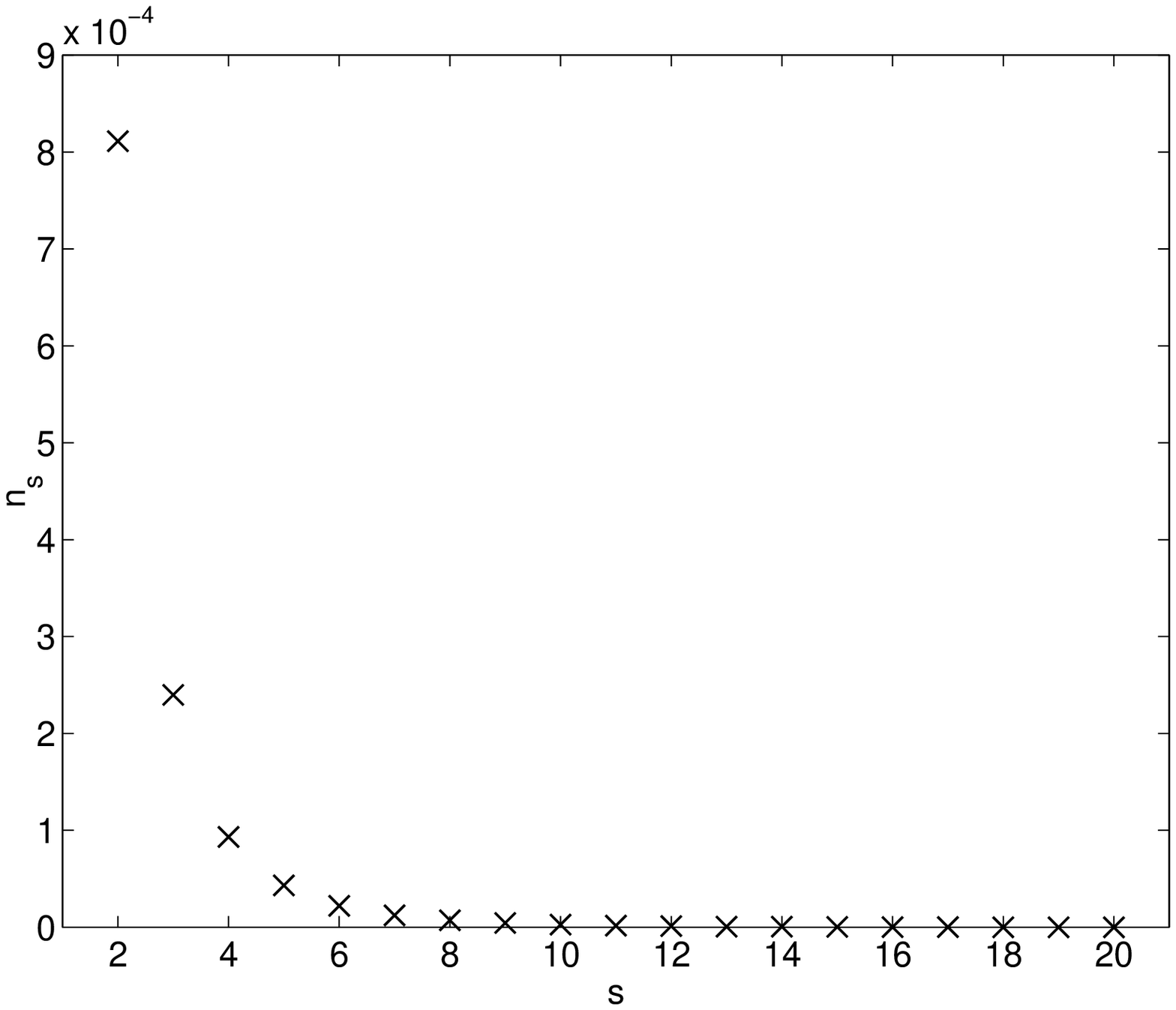,width=0.7\textwidth}} 
\vspace*{13pt} \fcaption{Cluster size distribution for
$p=0.5$.}\label{fig:s3}
\end{figure}

Cluster size distribution is surprising. In contrast with
Bernoulli percolation one can see the oscillations (see
Fig.~\ref{fig:s1}--\ref{fig:s2}). If $p=0.5$ (disordered medium)
then the distribution demonstrates classical behavior
(Fig.~\ref{fig:s3}). In Section~4 we will discuss a possible
connection such oscillations with the experiments and the real
world systems.

\section{Numerical estimation of percolation probabilities}
\begin{figure}[htbp]
\vspace*{13pt}
\centerline{\psfig{file=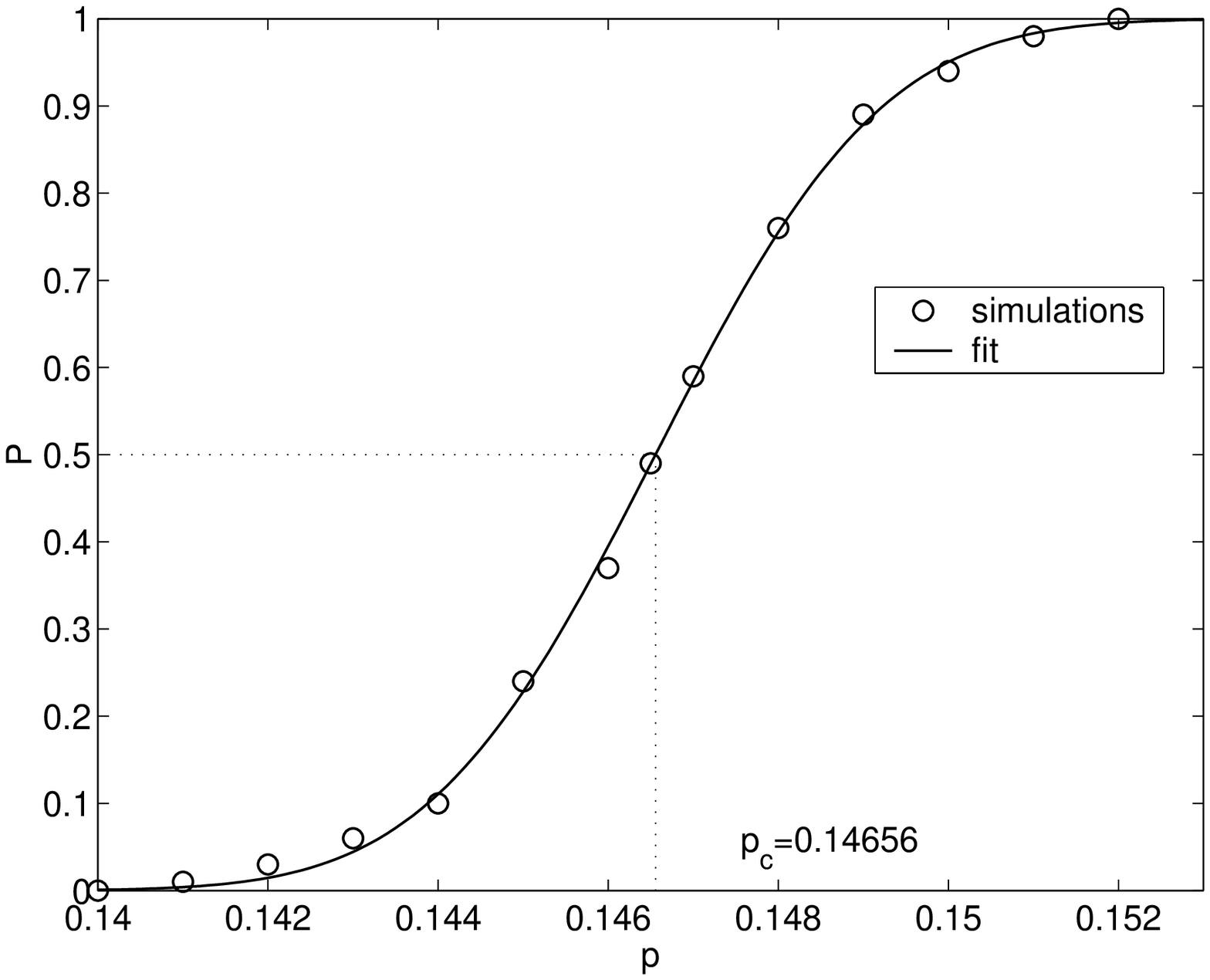,width=0.7\textwidth}} 
\vspace*{13pt} \fcaption{Percolation frequency as a function of
$p$ ($L=128$) (high resolution).}\label{fig:threshb}
\end{figure}

\noindent Estimates for $p_c$ have been obtained by means of
percolation frequencies. Simulations give the percolation
frequencies, which serve as an approximation of the percolation
probability. Critical percolation have been estimated by
not-linear fit functions defined by
\begin{equation}
p =\frac 12 (1 + {\rm erf} ( (p - p_c)/a)) \label{eq:fit}
\end{equation}
and
\begin{equation}
p =\left(1 + \exp \left( -(p - p_c)a\right)\right)^{-1}.
\label{eq:fit1}
\end{equation}
The first function converges to a step function as $a \to 0$ and
the second one as $a \to \infty$. Moreover we used a polynomial of
3-th degree. The fits gave almost the same results for the places
(the values of $p$) where the spanning probability curve reaches a
value of $1/2$. The similar approach to estimate the critical
probabilities was performed in works.\cite{Reimann,Babalievski}
Numerical estimation suggests that $p_c \approx 0.147$ for $L=128$
(Fig.~\ref{fig:threshb}).

\begin{figure}[htbp]
\vspace*{13pt}
\centerline{\psfig{file=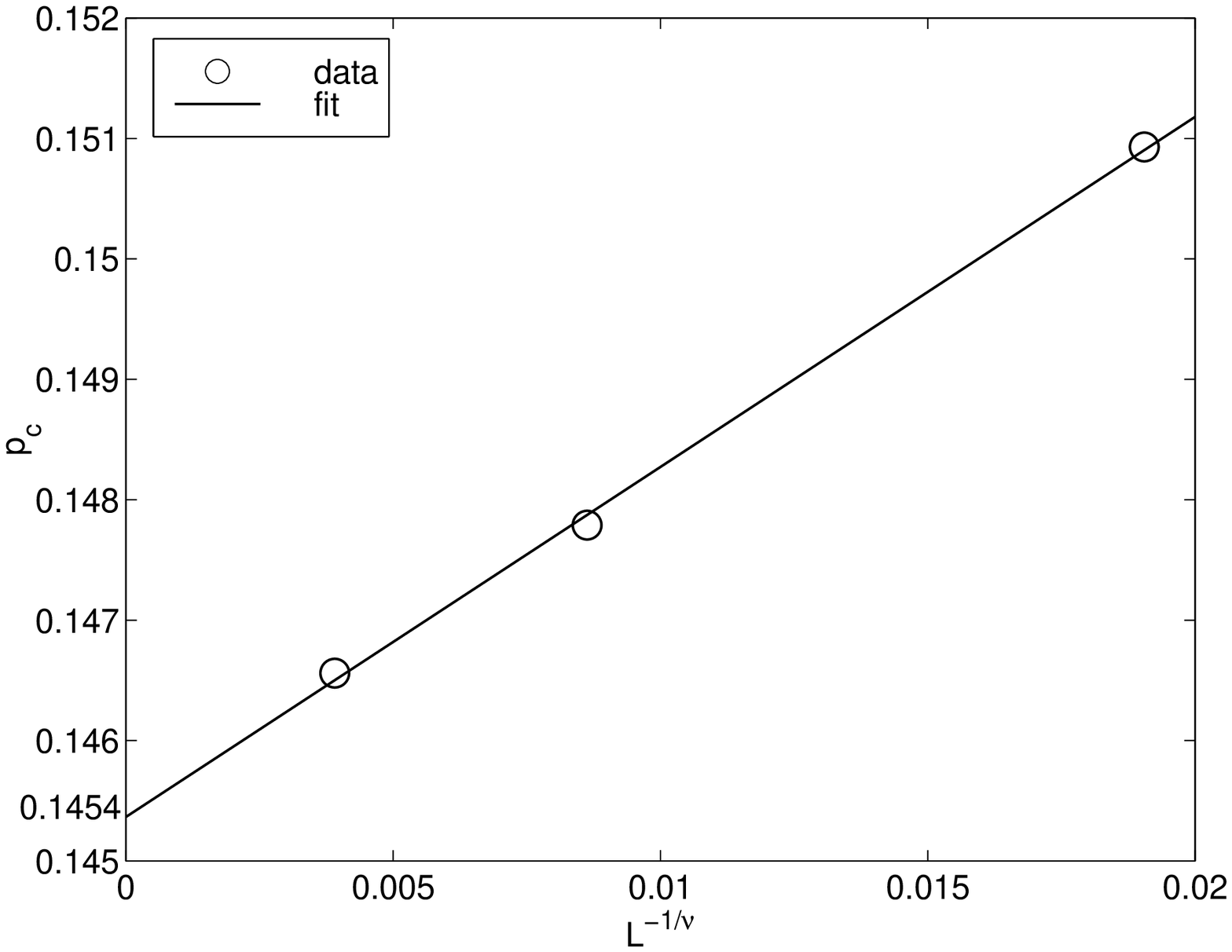,width=0.7\textwidth}} 
\vspace*{13pt} \fcaption{The percolation threshold as a function
of $L^{-1/\nu}$ .} \label{fig:scal}
\end{figure}

The percolation threshold was calculated for three values of the
linear lattice size $L = 32, 64, 128$. The percolation threshold
$p_c(\infty)$ for infinite lattices can be found by fitting these
results for different lattice sizes to the scaling
relation\cite{Stauffer}
\begin{equation}
   \left| p_c(L) - p_c(\infty) \right|  \propto  L^{-1/\nu},
   \label{eq:scal}
\end{equation}
where the critical exponent $\nu$ has the value $0.875$ in three
dimensions.\cite{Bunde} This method leads to an estimate
$p_c(\infty) \approx  0.1454$ (Fig.~\ref{fig:scal}).

We observed some infinity clusters at percolation threshold,
nevertheless we cannot assert there is not a finite size
effect.\cite{Stauffer97}

\section{Possible applications to the double 1:1 perovskites}%
\begin{figure}[htbp]
\vspace*{13pt}
\centerline{\psfig{file=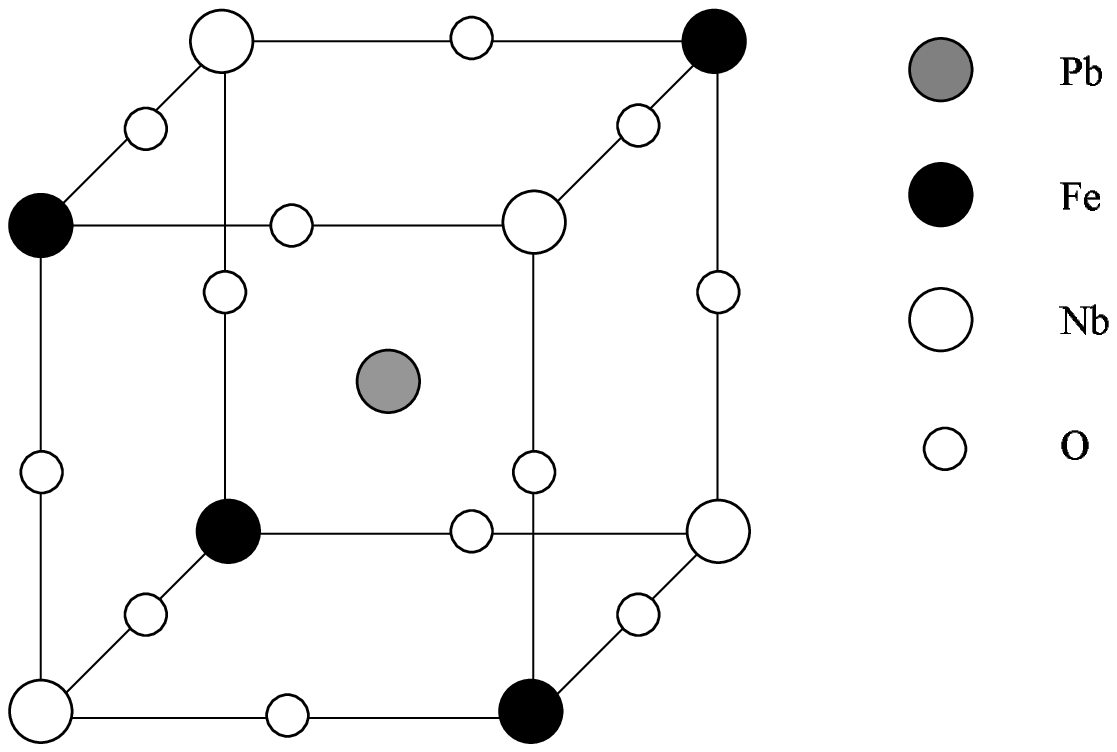,width=0.7\textwidth}} 
\vspace*{13pt} \fcaption{An example of ordered 1:1 double
perovskite: PbFe$_{1/2}$Nb$_{1/2}$O$_3$. Atoms of Fe and Nb form
an alternating simple cubic lattice.} \label{fig:pbfenbo3}
\end{figure}

\label{sec:appl}%
\noindent In last decades, much attention has been paid to ABO$_3$
perovskite-type oxides due to their unique dielectric,
electrooptic, and other properties. Complex perovskite-like oxides
of the AB$_x$B$'_{1-x}$O$_3$ type, in particularly double 1:1
perovskites {AB$_{1/2}$B$_{1/2}'$O$_3$}\footnote{Such the alloys
are denoted as A$_2$BB'O$_6$, too.}, e.g.
(Ba,\,Sr)Fe$_{1/2}$Mo$_{1/2}$O$_{3}$ and PbFe$_{1/2}$Nb$_{1/2
}$O$_{3}$, have much more surprising and useful properties.
Examples of their applications are piezoelectric transducers and
actuators, non-volatile ferroelectric memories, and
microelectronic devices. The structure is built up by ordering
perovskite blocks in a rock salt superlattice
(Fig.~\ref{fig:pbfenbo3}) and the properties of the material are
thought to critically depend on this ordering. The double
perovskite Sr$_2$FeMoO$_6$ and related materials are good
candidates for magnetic devices, as they combine a high Curie
temperature and a fully polarized (half metallic) conduction band.
For more detailed overview and a large collection of references
see e.g.\cite{Balcells,Alonso}.

Little thing is known and understood about the microscopic
mechanisms responsible for their very convenient properties. At
the present, these materials are being extensively studied. One
particular aspect that seems very promising  is the effect of
atomic ordering on the properties of these alloys. Anti-site
disordering in Sr$_2$FeMoO$_6$ double perovskites (containing B
atoms at B$'$ positions, and vice versa) has recently been shown
to have a dramatic influence in their magnetic and
magnetotransport properties. Experiments suggest that, in many
samples, the saturation magnetization is less than the expected
value. This effect is usually ascribed to the presence of antisite
defects, where, due to the similarity of their atomic radii, Mo
ions are randomly placed on the Fe sublattice and conversely.

A model described above seems very promising to understand some
features of the double perovskites. In particularly, the
oscillations of the hyperfine magnetic fields distribution in the
PbFe$_{1/2}$Nb$_{1/2}$O$_3$ samples\cite{Raevskii} may be related
with the oscillations of the cluster size distribution
(Fig.~\ref{fig:s1},\ref{fig:s2}).

\nonumsection{Acknowledgements} \noindent The authors are grateful
to Prof. R.~V.~Vedrinskii for pointing out the problem. They thank
Prof. S.~A.~Prosandeev and Prof. I.~P.~Raevskii for discussions.

\nonumsection{References}

\end{document}